\documentclass[aps, pra, longbibliography,superscriptaddress]{revtex4-1}
\usepackage[english]{babel}
\normalsize 

\usepackage{amsmath,amsfonts}
\usepackage{rotating}
\usepackage{leftidx}
\usepackage{here}
\usepackage{pbox}
\usepackage[braket,qm]{qcircuit}
\usepackage[nolist,nohyperlinks]{acronym}
\usepackage{physics}

\usepackage{float}
\usepackage[dvipsnames]{xcolor}
\usepackage{graphicx}
\usepackage{caption}
\usepackage{subcaption}
\usepackage{tikz}
\usetikzlibrary{external}
\usetikzlibrary{decorations.pathreplacing,calligraphy,positioning,shapes.misc}
\usepackage{pgfplots}
\usepackage{natbib}
\usepackage{dcolumn} 
\usepackage{multirow}
\usepackage{booktabs}

\usepackage[normalem]{ulem}

\newcommand{\fett}[1]{\mbox{\boldmath$#1$}}

\newcommand{\RR}{ \fett R}
\newcommand{\acr}[1]{\hat {\textmd a}_{#1}^{\dagger}}
\newcommand{\aan}[1]{\hat {\textmd a}_{#1}}

\renewcommand{\vec}[1]{\boldsymbol{#1}}

\definecolor{tabblue}{HTML}{1f77b4}
\definecolor{taborange}{HTML}{ff7f0e}
\definecolor{tabgreen}{HTML}{2ca02c}
\pgfplotsset{compat=1.18}
\pgfplotsset{grid style={line width=0.1pt, color=black!5}}

\newcommand{\plotwidth}{0.8\textwidth}

\begin{document}

\title{Nonadiabatic nuclear-electron  
dynamics: a quantum computing approach}

\author{Arseny Kovyrshin} 
\email{arseny.kovyrshin@astrazeneca.com}
\affiliation{ Data Science and Modelling, Pharmaceutical Sciences, R\&D, AstraZeneca Gothenburg, Pepparedsleden 1, Molndal SE-431 83, Sweden}
\author{M{\aa}rten Skogh} 
\email{marten.skogh@astrazeneca.com}
\affiliation{ Data Science and Modelling, Pharmaceutical Sciences, R\&D, AstraZeneca Gothenburg, Pepparedsleden 1, Molndal SE-431 83, Sweden}
\affiliation{ Department of Chemistry and Chemical Engineering, Chalmers University of Technology, Gothenburg, Sweden} 
\author{Lars Tornberg} 
\email{lars.tornberg@astrazeneca.com}
\affiliation{ Data Science and Modelling, Pharmaceutical Sciences, R\&D, AstraZeneca Gothenburg, Pepparedsleden 1, Molndal SE-431 83, Sweden}
\author{Anders Broo}
\email{anders.broo@astrazeneca.com} 
\affiliation{ Data Science and Modelling, Pharmaceutical Sciences, R\&D, AstraZeneca Gothenburg, Pepparedsleden 1, Molndal SE-431 83, Sweden}

\author{Stefano Mensa}
\email{stefano.mensa@stfc.ac.uk}
\author{Emre Sahin}
\author{Benjamin C. B. Symons}
\email{benjamin.symons@stfc.ac.uk}
\affiliation{The Hartree Centre, STFC, Sci-Tech Daresbury, Warrington, WA4 4AD, United Kingdom}
\author{Jason Crain}
\email{jason.crain@ibm.com}
\affiliation{IBM Research Europe, Hartree Centre STFC Laboratory, Sci-Tech Daresbury, Warrington WA4 4AD, United Kingdom}
\affiliation{Department of Biochemistry, University of Oxford, Oxford, OX1 3QU, UK}
\author{Ivano Tavernelli}
\email{ita@zurich.ibm.com}
\affiliation{IBM Quantum, IBM Research Europe – Zurich, S\"aumerstrasse 4, 8803 R\"uschlikon, Switzerland}

\begin{abstract}

The combined quantum electron-nuclear dynamics is often associated with the Born--Huang expansion of the molecular wave function and the appearance of nonadiabatic effects as a perturbation. On the other hand, native multicomponent representations of electrons and nuclei also exist, which do not rely on any \textit{a priori} approximation. However, their implementation is hampered by prohibitive scaling costs and therefore quantum computers offer a unique opportunity for extending their use to larger systems. Here, we propose a quantum algorithm for the simulation of the time-evolution of molecular systems in the second quantization framework, which is applied to the simulation of the proton transfer dynamics in malonaldehyde.  After partitioning the dynamics into slow and fast components, we show how the entanglement between the electronic and nuclear degrees of freedom can persist over long times if electrons are not adiabatically following the nuclear displacement. The proposed quantum algorithm may become a valid candidate for the study of electron-nuclear quantum phenomena when sufficiently powerful quantum computers become available. 

\end{abstract}
\maketitle

\begin{acronym}
\acro{vrte}[VRTE]{Variational Real-Time Evolution}
\acro{vite}[VITE]{Variational Imaginary-Time Evolution}
\acro{nisq}[NISQ]{Noisy Intermediate-Scale Quantum}
\acro{vha}[VHA]{Variational Hamiltonian Ansatz}
\acro{bo}[BO]{Born--Oppenheimer}
\acro{pbo}[pBO]{pre-Born--Oppenheimer}
\acro{nomo}[NO+MO]{Nuclear Orbital plus Molecular Orbital}
\acro{neo}[NEO]{Nuclear-Electronic Orbital}
\acro{neohf}[NEOHF]{Nuclear Electronic Orbitals Hartree--Fock}
\acro{hf}[HF]{Hartree--Fock}
\acro{uhf}[UHF]{Unrestricted Hartree--Fock}
\acro{sto}[STO-3G]{single-$\zeta$ (minimal) basis set contracted from 3 Gaussian primitives}
\acro{sto6}[STO-6G]{single-$\zeta$ (minimal) basis set contracted from 6 Gaussian primitives}
\acro{631}[6-31G]{split-valence double-$\zeta$ Gaussian basis set in 6-31 contraction scheme}
\acro{dzsnb}[DZSNB]{split-valence double-$\zeta$ nuclear basis set composed of 2 uncontracted Cartesian $S$ functions}
\acro{vqe}[VQE]{Variational Quantum Eigensolver}
\acro{avqe}[AdaptVQE]{Adaptive Variational Quantum Eigensolver}
\acro{neocasci}[NEOCASCI]{Nuclear-Electronic Orbital Complete Active Space Configuration Interaction}
\acro{casci}[CASCI]{Complete Active Space Configuration Interaction}
\acro{neofci}[NEOFCI]{Nuclear-Electronic Orbital Full Configuration Interaction}
\acro{ci}[CI]{Configuration Interaction}
\acro{fci}[FCI]{Full Configuration Interaction}
\acro{ucc}[UCC]{Unitary Coupled Cluster}
\acro{neoucc}[NEOUCC]{Nuclear-Electronic Orbitals Unitary Coupled Cluster}
\acro{uccsd}[UCCSD]{Unitary Coupled Cluster Singles and Doubles}
\acro{neouccsd}[NEOUCCSD]{Nuclear-Electronic Orbitals Unitary Coupled Cluster Singles and Doubles}
\acro{neouccsdt}[NEOUCCSDT]{Nuclear-Electronic Orbitals Unitary Coupled Cluster Singles, Doubles, and Triples}
\acro{lih}[Li-H]{lithium hydride}
\acro{h2}[H$_2$]{hydrogen molecule}
\acro{c3h5o2}[C$_3$H$_5$O$_2$]{protonated enol malonaldehyde form}
\acro{qpe}[QPE]{Quantum Phase Estimation}
\acro{cobyla}[COBYLA]{Constrained Optimization By Linear Approximation}
\acro{cg}[CG]{Conjugate Gradient}
\acro{slsqp}[SLSQP]{Sequential Least SQuares Programming}
\acro{mp2}[MP2]{the second-order M\o{}ller--Plesset}
\acro{ts}[TS]{transition state}
\acro{on}[ON]{Occupation Number}
\end{acronym}

\newpage

\bigskip

\textit{Introduction.} 
Most strategies for practical simulation of materials at the molecular scale make two fundamental assumptions: firstly, that atomic nuclei behave classically  and, secondly, that electronic and nuclear dynamics are adiabatically separable (the Born--Oppenheimer (BO) approximation) such that the system can be described by a product of stationary eigenstates. 
Both can fail under certain important circumstances. 
In particular, the BO approximation requires that time-dependent perturbations preserve the instantaneous electronic ground state --- a condition satisfied only if the perturbation is sufficiently slow and the energy separation between the ground state and other low-lying levels is sufficiently large.  If, however, the timescales of electronic and nuclear degrees of freedom become closer or the energy gap separating potential energy surfaces becomes small (as it does near avoided crossings or conical intersections)  nonadiabatic effects arise. Such situations occur when, for example, electron dynamics are driven by ultrafast laser pulses or excited state reaction pathways traverse quasi-degenerate electronic levels. 
Also, despite the small de Broglie wavelengths of nuclei relative to electrons,
nuclear delocalization still occurs over length scales that are important in chemical processes. In these regimes, the classical nuclei approximation also fails.

Consequently, reactions involving light elements can be influenced significantly by quantum effects, \textit{e.g.}, tunneling, whereby the reaction takes a route through a classically forbidden region in configuration space (`through' the energy barrier) in contrast to conventional transition state theory~\cite{wild2023rate,schreiner2008capture}.
In some cases, non-classical processes can even supersede traditional kinetics, driving  reactions exclusively toward a product for which the classical path would have a higher barrier~\cite{schreiner2011methylhydroxycarbene}. In biological systems, quantum phenomena involving light nuclei  are known to influence catalysis and enzymatic activity~\cite{klinman2013hydrogen,cha1989hydrogen} with implications, \textit{e.g.},  in the design of novel inhibitors~\cite{nagel200921st}.  In particular, the observed rates of certain enzyme-catalysed reactions cannot be accounted for without introducing corrections for proton tunnelling~\cite{bavnacky1981dynamics,basran1999enzymatic}. 
Also, in DNA chemistry, there is evidence pointing towards proton tunnelling playing a role in the generation of abnormal base pairs (tautomers), which may be implicated in a particular class of mutations~\cite{slocombe2022open}. 

Significant and fundamental practical challenges limit the extent to which  non-classical features of both electrons and nuclei  can be incorporated together into high-fidelity molecular simulations. 
It is well known that methods to solve the full Schr\"odinger equation exactly on a classical computer exhibit prohibitive scaling. 
In fact, the dimension of the Hilbert space increases exponentially with the system size while the complexity of determining solutions scales factorially with the number of basis functions~\cite{verma2021scaling}.  
This is further exacerbated when the descriptions are extended to the treatment of coupled electron-nuclear dynamics.  

However, quantum computation has opened new prospects for improving the scaling behaviour in this important class of problem. 
In a recent publication~\cite{kovy2023} we introduced a new quantum computing algorithm, based upon the \ac{neo} framework,~\cite{webb2002} for the efficient  treatment of quantum electron-nuclear effects on near-term quantum computers. 
In particular, we demonstrated how the \ac{neo} approach can be applied to the study of multi-component quantum mechanical systems composed of electrons and a selected set of light nuclei (protons) in order to go beyond the Born–Oppenheimer approximation~\cite{zhao2020real,pavošević2020multicomponent}.

\begin{figure}[ht]
    \centering
    \includegraphics[width=0.57\textwidth]{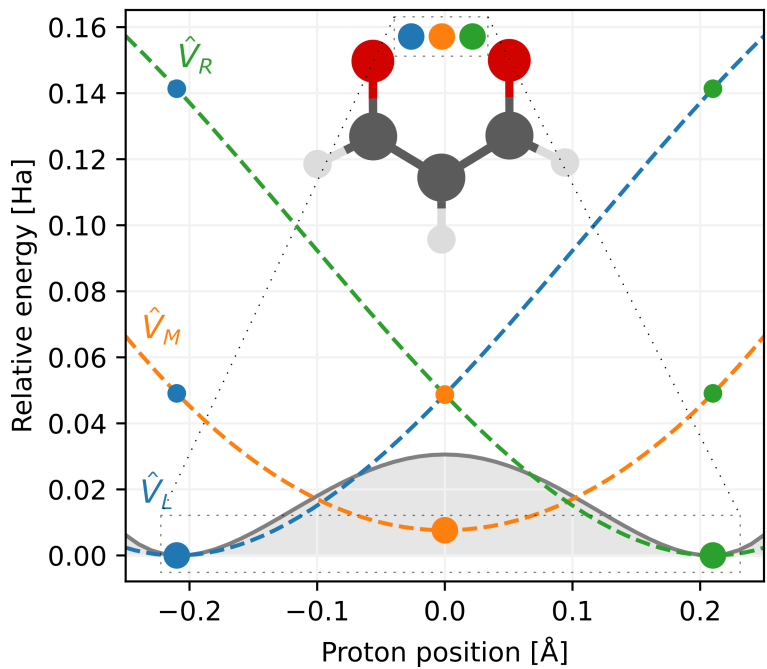}
    \caption{A schematic image of the potentials from the three Hamiltonians ($R$, $L$, and $M$) used throughout this work. The inset malonaldehyde shows the transition state structure with `Left', `Middle', and `Right' settings for proton. Apart from the discrete points shown, the potential energy curves are not exact and merely serve to provide the reader with a general guide. The curves were generated by fitting a third-order polynomial to the three points corresponding to each Hamiltonian. The grey shaded area is the scaled-up electronic potential energy calculated with MP2, and aims to illustrate the total potential, and in particular the reaction barrier.}
    \label{fig:schematic}
\end{figure}

In this work, we extend the framework to study the reaction dynamics of chemical systems with important electron-nuclear quantum effects. To this end, we consider malonaldehyde as a model system for proton transfer involving intramolecular hydrogen bonds~\cite{coutinho2004ground}. 
The key structural feature is the `O$-$H$\cdots$O' hydrogen bond (see Fig.~\ref{fig:schematic}) for which there are two possible asymmetric configurations leading to a double-well potential with a proton barrier separating the isomers.
This system has previously been investigated using the \ac{neo} framework coupled with density functional theory on classical processors ~\cite{yu2022nonadiabatic}.
Here, the aim is to demonstrate real-time evolution of the proton transfer process using a quantum computing implementation designed to exploit quantum speed-up. 

\textit{Theory.}
In the \ac{neo} approach~\cite{webb2002} we make use of Slater determinants comprised of both electronic and protonic spin orbitals. These are expanded with separate nuclear and electronic Gaussian basis sets~\cite{webb2002} and optimized using the \ac{neohf} method. Note that, by analogy to the alpha and beta electrons in the \ac{uhf} approach, electrons and protons in \ac{neohf} only interact through the Coulomb potential. 
We can then construct the corresponding 
multi-particle basis set for the \ac{neofci} approach~\cite{webb2002}, leading to the following second quantization representation of the electron-nuclear Hamiltonian~\cite{webb2002}
\begin{align}
\hat H = & 
\sum_{ij} h_{ij}\, \acr{i} \aan{j}
+\sum_{IJ} h_{IJ}\, \acr{I} \aan{J} 
+\frac{1}{2}\sum_{ijkl} h_{ijkl}\, \acr{i} \acr{k} \aan{l} \aan{j}
+ \frac{1}{2}\sum_{IJKL} h_{IJKL}\, \acr{I} \acr{K} \aan{L} \aan{J} \nonumber\\
&-\sum_{ijKL} h_{ijKL}\, \acr{i} \acr{K} \aan{L} \aan{j} 
+\sum_{IJ,A} h_{IJ,A}\, \acr{I} \aan{J}
-\sum_{ij,A} h_{ij,A}\, \acr{i} \aan{j}
+\frac{1}{2}\sum_{AB} \frac{Z_A Z_B}{|\RR_A-\RR_B|}.
\label{eq:neofHam}
\end{align}
Here $h_{pq}$ and $h_{pq, A}$ are one-particle integrals involving kinetic energy and interactions with classical nuclei, respectively; $h_{pqrs}$ are two-particle integrals responsible for interactions between quantum particles; $\acr{}$ and $\aan{}$ are the fermionic creation and annihilation operators; $Z$ is nuclear charge; and lastly, $\RR$ is the nuclear coordinate for classical, point-like nuclei (for details see Ref.~\cite{kovy2023}). In Eq.~\eqref{eq:neofHam} upper case indices label protonic spin orbitals while lower case indices are used for electronic spin orbitals. Indices $A$ and $B$ denote classical nuclei, while $I, J, K$, and $L$ are reserved for nuclei treated quantum mechanically.
Note that creation and annihilation operators must obey anti-commutation relations for indistinguishable fermions (proton-proton or electron-electron) and commutation relations between distinguishable fermions (proton-electron). 

As we aim to study chemical reaction dynamics, the \ac{neo} wave function acquires time dependence through the time-dependent \ac{ci} coefficients, $C_{\mu \nu}(t)$. Accordingly, the time-dependent nuclear-electronic wave function is given by
\begin{align}\label{eq:neofci}
\ket{\Psi(t)} = \sum_{\mu \nu}C_{\mu \nu}(t) \ket{\Phi_{\mu}^e} \ket{\Phi_{\nu}^n} \, ,
\end{align}
where  $\ket{\Phi_{\mu}^e}$ and $\ket{\Phi_{\nu}^n}$ are electronic and nuclear configurations, respectively.
It is important to stress that neither the Slater determinants nor the molecular orbitals are allowed to change during the dynamics,  but only the \ac{ci} coefficients $C_{\mu \nu}(t)$ evolve in time according to the time-dependent Schr\"odinger equation for the molecular Hamiltonian in Eq.~\eqref{eq:neofHam}. Thus, the evolution of the wave function can be expressed by the equation of motion for \ac{ci} coefficients
\begin{equation}
\vec{H} \vec{C} = i\frac{\partial}{\partial t}\vec{C}\, ,
\end{equation}
where matrix elements of $\vec{H}$ are defined through
\begin{equation}
H_{\kappa \lambda, \mu \nu}=\bra{\Phi^e_{\kappa}}\bra{\Phi^n_{\lambda}} \hat H \ket{\Phi^e_{\mu}} \ket{\Phi^n_{\nu}}.
\end{equation}
On a quantum computer, the initial \ac{neo} wave function, $\ket{\Psi(t_0)}$, will be efficiently approximated using the \ac{neouccsdt} ansatz~\cite{kovy2023}.

As a model for the demonstration of our quantum algorithm for electron-proton quantum dynamics, we consider the proton transfer process in malonaldehyde using the setup discussed in our previous publication~\cite{kovy2023}.
As shown in Fig.~\ref{fig:schematic}, the proton can be localized at two possible asymmetric equilibrium positions separated from each other by 0.42 {\AA} (marked with blue and green spheres) as well as at the peak of the barrier in the middle (orange sphere). 
The two asymmetric structures are characterized by a different relaxation of the electronic orbitals and are associated with the formation of two different OH bonds: one with the oxygen on the right (green sphere) and the other with the oxygen on the left (blue sphere) in the inset of Fig.~\ref{fig:schematic}. 
Finally, the symmetric setting has the proton shared equally between the two oxygen atoms. 
These setups define three different Hamiltonians: $\hat H_R$ ($R$ for `Right'),  $\hat H_L$ ($L$ for `Left'), and $\hat H_M$ ($M$ for `Middle'), respectively. The nuclear and electronic orbitals were optimized with \ac{neohf} calculations using the transition state molecular scaffold of malonaldehyde~\cite{kovy2023} and imposing $C_s$ symmetry. For protons, a \ac{dzsnb} was used. Electronic orbitals were represented using a \ac{631}. 
In the asymmetric cases described by $\hat H_R$  and $\hat H_L$, the \ac{neohf} calculations were performed with two nuclear basis functions at the equilibrium position (green and blue spheres respectively). 
Whereas, in the symmetric case, the two nuclear basis functions were located at the barrier's maximum (orange sphere). 
In all cases, only the transferred proton was treated quantum mechanically (described by two nuclear orbitals placed at one of the locations), while all other nuclei were considered as classical point charges at fixed positions. We also kept the same set of electronic basis functions in all three cases, putting electronic basis functions at all proton locations ($R$, $L$, and $M$). 

\begin{figure}[ht]
\includegraphics[]{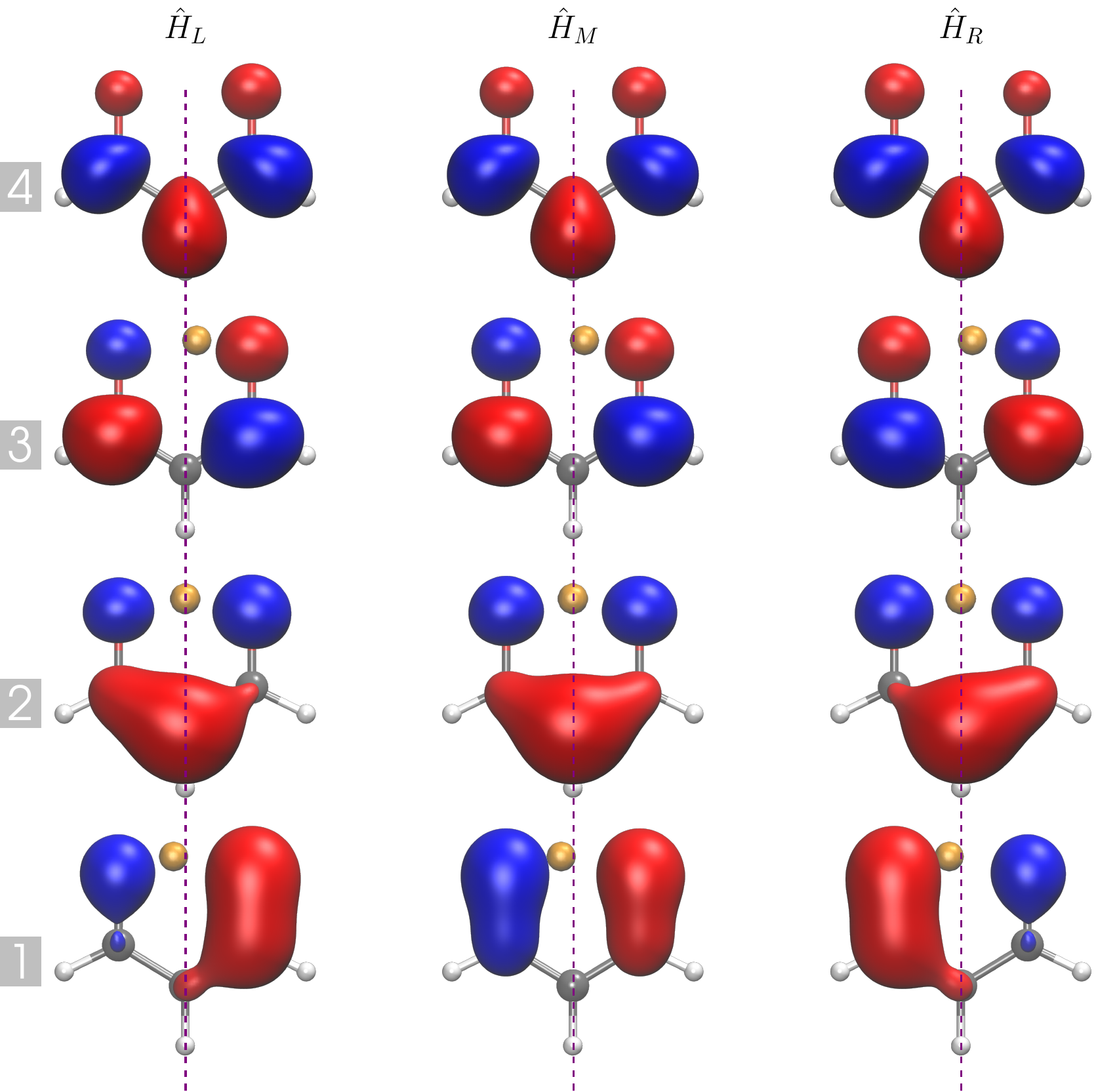}
    \centering
    \caption{Electronic (red/blue) and nuclear (orange) orbitals (isosurface value is $0.05$) included in the active space of \ac{neocasci} calculation for malonaldehyde in the ``Left'' ($\hat H_L$), ``Middle'' ($\hat H_M$), and ``Right'' ($\hat H_R$)  setups. The orbitals have been prepared with a \ac{neohf} calculation using the \ac{dzsnb} and \ac{631} basis sets for the nuclei and the electrons, respectively.}
    \label{fig:malon_orbitals}
\end{figure}

For each of the three sets of electronic orbitals (corresponding to the Hamiltonians $\hat H_R$, $\hat H_M$, and $\hat H_L$) we performed \ac{neocasci} calculations to obtain the corresponding reference energies, see Table~\ref{tab:Malon}. To keep the electronic active space compact, in all calculations the 17 low-lying core orbitals for electrons were considered fully occupied and only 4 orbitals hosting 4 electrons were considered active. 
The nuclear active space consisted of the lowest-energy nuclear orbitals from the $\hat H_R$, $\hat H_M$, and $\hat H_L$ \ac{neohf} setups. 
Thus, the nuclear active space for all three \ac{neocasci} calculations remains the same and consists of 3 orbitals.  
Electronic and nuclear \ac{neohf} orbitals used in the active space for all three settings are shown in Figure~\ref{fig:malon_orbitals}. 
Based on the \ac{neocasci} calculations at these three setups, the barrier is estimated to be 0.0051 Hartree. This is rather close to the value obtained in Ref.~\cite{kovy2023}, 0.0050 Hartree.

To enable calculations on a quantum computer, we mapped the second-quantized Hamiltonians ($\hat H_R$, $\hat H_M$, and $\hat H_L$) spanning 14 spin orbitals to corresponding 8-qubit Hamiltonians, with 6 qubits spanning the electronic subspace, and 2 qubits spanning the nuclear one. The parity fermion-to-qubit transformation  for electronic and nuclear operators was used together with the qubit reduction techniques as described  in Ref.~\cite{kovy2023}. 
The energy values for all systems estimated with the \ac{vqe} and the \acs{neouccsdt} ansatz are presented in Table~\ref{tab:Malon}. 
The \ac{neouccsdt} results are in agreement with the \ac{neocasci} references for all settings, 
and the value for the energy barrier agrees with the reference value to within $1\times10^{-6}$ Hartree, which confirms the excellent quality of the present setup for studying proton dynamics. 
We also report the von Neumann entropy, $s$, capturing the entanglement between the nuclear and electronic subsystems. This is obtained according to
\begin{equation}\label{eq:vonNeuman}
s=-{\rm Tr}( \hat \rho_{e} \ln \hat \rho_{e}) = -{\rm Tr}( \hat \rho_{n} \ln \hat \rho_{n}) \,,
\end{equation}
where $\hat \rho_{e}={\rm Tr}_{n} \left[\hat \rho_{e,n}\right]$ and $\hat \rho_{n}={\rm Tr}_{e} \left[\hat \rho_{e,n}\right]$ are the reduced density matrices corresponding to electrons and nuclei respectively, and $\hat \rho_{e,n}$ is the full density matrix.
In contrast to our previous study~\cite{kovy2023}, the proton-electron entanglement in the $\hat H_L$ and $\hat H_R$ ground states is no longer zero but amounts to 0.0020 and 0.0019 for \ac{neouccsdt} and \ac{neocasci}, respectively.  
This is mainly due to the extension of nuclear active space for $\hat H_L$ and $\hat H_R$ (compared to Ref.~\cite{kovy2023}), which includes orbitals at the top of, as well as on both sides of the barrier separating the two minima. 
Our calculations still show increasing electron-nuclear entanglement as the proton approaches the top of the barrier.

\begin{table}[htb]
\caption{Energies and von Neumann entanglement entropies, Eq.~\eqref{eq:vonNeuman}, for malonaldehyde obtained with \acs{neocasci} and \acs{neouccsdt} using $\hat H_L$, $\hat H_R$, and $\hat H_R$. The barrier height, $\Delta E$, is evaluated as the difference in energy between the L and M setups. We observe good agreement with the barrier height reported in Ref.~\cite{kovy2023} for \acs{neouccsdt}.}
\label{tab:Malon}
\begin{center}
\begin{tabular}{@{}llrrrrr@{}}
\toprule
 &            &  \multicolumn{2}{c}{$\hat H_L$/$\hat H_R$}   &\multicolumn{2}{c}{$\hat H_M$}& \\
\midrule
 & Method              & Energy/Ha & Entropy& Energy/Ha & Entropy & $\Delta$ E/Ha \\
\midrule
 &\ac{neocasci}~\cite{kovy2023}& -265.490948 & 0.0000 & -265.485937 & 0.0044 & 0.005011 \\
 &\ac{neocasci}& -265.491028 & 0.0020 & -265.485912 & 0.0038& 0.005116 \\
\midrule
\\
\\
 &\ac{neouccsdt}~\cite{kovy2023} & -265.490943 & 0.0000 & -265.485936 & 0.0044 & 0.005007 \\
 &\ac{neouccsdt} &-265.491023 & 0.0019 & -265.485909 & 0.0038 & 0.005115 \\
\midrule
\end{tabular}
\end{center}
\end{table}

\textit{Quantum dynamics.}
After preparing the ground state wave function for a given nuclear configuration, we can propagate it from the initial time $t_0$ to the final time $t_f$ by means of the time-evolution operator $\mathcal{U}(t_f,t_0)$ 
\begin{equation}\label{eq:time-ev}
    \ket{\Psi(t_f)} =  \mathcal{U}(t_f,t_0) \ket{\Psi(t_0)}\, . 
\end{equation}

\begin{figure}[htb]
\includegraphics[]{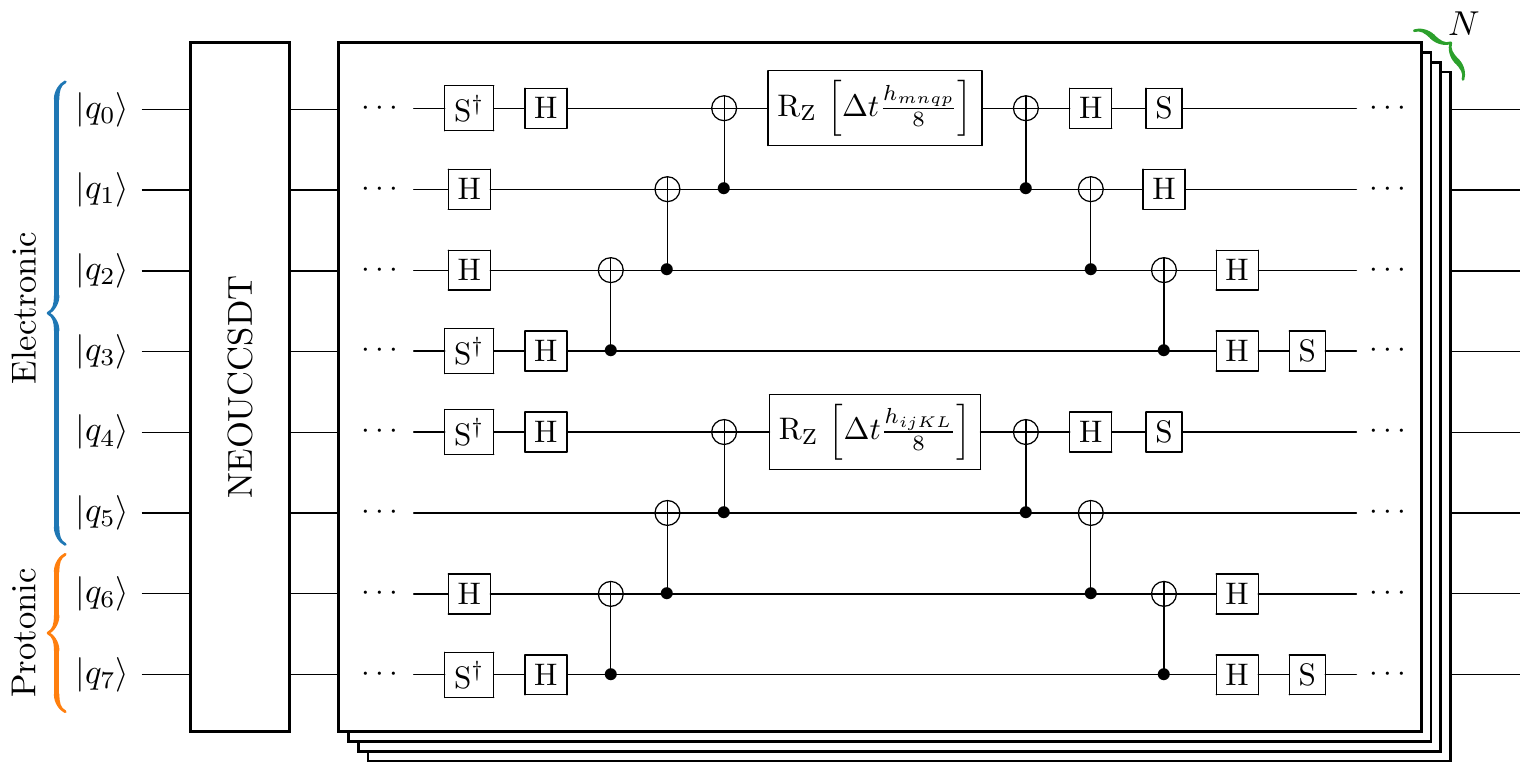}
    \caption{Schematic drawing of the time evolution operator, Eq.~\eqref{eq:time_ev_final}, implemented as a quantum circuit with a series of $N$ time steps of size $\Delta t$ on 8 qubits.
    }
    \label{fig:time_evo_circuit}
\end{figure}

The form of $U(t_f,t_0)$ depends on the Hamiltonian describing the system. For a general closed system, described by a time-dependent Hamiltonian,
the exact time evolution is given by the Dyson series
\begin{equation}\label{eq:dyson}
    \mathcal{U}(t_f,t_0) = \mathcal{T}\exp\left[ -i \int_{t_0}^{t_f} \hat{H}(t) dt\right],
\end{equation}
where $\mathcal{T}$ is the time-ordering operator. 
Expressing the time-dependent Hamiltonian, $\hat{H}(t)$, as a sum of weighted Pauli strings, $P_k$, with time-dependent weights $h_{k}(t)$
\begin{equation}
    \hat{H}(t) = \sum_{k=1}^{K} h_{k}(t) P_{k},
\end{equation}
we can approximate time-ordered Dyson series with the first-order decomposition formula given by Suzuki~\cite{suzuki1993general}. Specifically, breaking the evolution into a series of $N$ discrete steps of size $\Delta t = t_f/N$, Eq.~\eqref{eq:dyson} can be approximated as follows
\begin{equation}
    \mathcal{U}_1(t,t_0) \approx \prod_{t_j=t_0}^{t_{N-1}} U(t_{j} + \Delta t ,t_j) = \prod_{t_j=t_0}^{t_{N-1}} \prod_{k=1}^{K} e^{-ih_{k}(t_j + \Delta t/2) P_{k}\Delta t}.
\label{eq:time_ev_final}
\end{equation}
The corresponding quantum circuit is shown schematically in Fig.~\ref{fig:time_evo_circuit}.

It should be noted that, with such an approach, the number of time steps required to achieve a desired time evolution typically results in circuits that are too deep to be implemented on noisy, near-term hardware. On the other hand, other noise-resilient time-propagation algorithms such as \ac{vrte}~\cite{yuan2019,mcar2019} are also available.

In this work, we consider time evolution driven by a linear combination of the three Hamiltonians described above: $\hat H_L$, $\hat H_M$, and $\hat H_R$. 
These Hamiltonians share the same pure nuclear contributions, specifically, the second, the sixth, and the eighth terms in Eq.~\eqref{eq:neofHam} (the fourth term is absent in our setup as only one proton is considered). The pure electronic and mixed electron-nuclear terms vary due to the different electronic active spaces employed, see Figure~\ref{fig:malon_orbitals}. 
Due to the large potential energy barrier separating the minima of $\hat H_L$ and $\hat H_R$, we cannot expect a spontaneous transition of the shared proton from one minimum to the other.
To capture such dynamics, we opted for a parameterized time-dependent Hamiltonian, made of a normalized linear combination of the three original Hamiltonians ($\hat H_L$, $\hat H_M$, and $\hat H_R$)
\begin{equation}
\label{eq:H_linear_comb}
\hat H(t) = \alpha(t) \hat H_L + \beta(t) \hat H_M + \gamma(t) \hat H_R, 
\end{equation}
where normalization requires that $\alpha(t) + \beta(t) + \gamma(t) = 1 $. 

\begin{figure}[ht]
\includegraphics[]{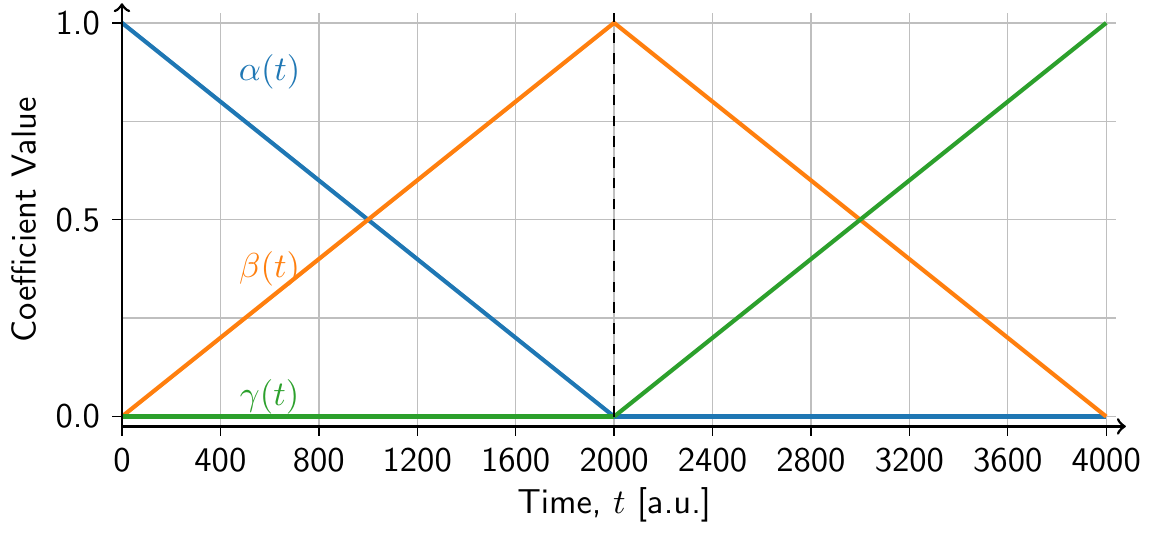}
    \caption{Adiabatic mixing of the Hamiltonians. The different Hamiltonians are mixed in pairs. Initial mixing is performed between $\hat H_L$ and $\hat H_M$  through $\alpha(t)\hat H_L + \beta(t)\hat H_M$, with the subsequent mixing of $\hat H_M$ and $\hat H_R$ as $\beta(t)\hat H_M + \gamma(t)\hat H_R$. }
    \label{fig:mixing_weights}
\end{figure}

After preparing the ground state of the Hamiltonian $\hat H_L$ ($\alpha(t_0)=1,\beta(t_0)=\gamma(t_0)=0$), $\ket{\Psi_L}$, we will `adiabatically’ drive the system towards the ground state of  $\hat H_R$ (namely, $\ket{\Psi_R}$) by updating the Hamiltonian's parameters in Eq.~\eqref{eq:H_linear_comb} following the profiles in Fig.~\ref{fig:mixing_weights}, and evolving the state of the system using Eq.~\eqref{eq:time_ev_final}. 
Given a final time, in this case $t_f=4000$ a.u., this protocol guarantees a smooth transition in the Hamiltonian space from $\hat H_L$ at $t=0$ to  $\hat H_R$ at $t=t_f$ passing through $\hat H_M$. 

\begin{figure}[ht]
    \centering
    \includegraphics[width=\plotwidth]{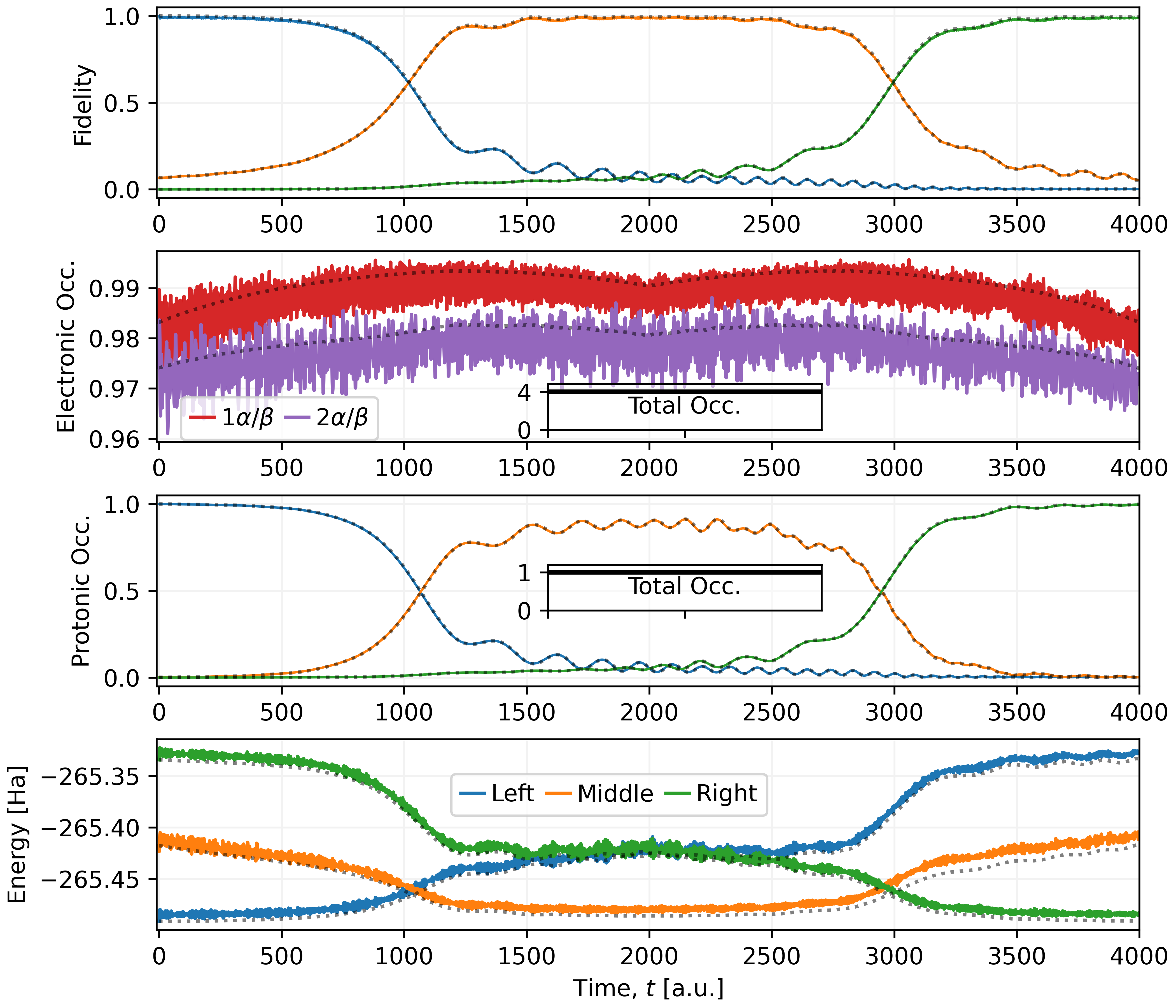}
    \caption{
    Adiabatic evolution performed for time $t_f=4000$ a.u. with time step $\Delta t =1.0$ a.u.. 
    The black dotted lines show the reference results, generated by 4th-order Runge--Kutta (RK4)~\cite{num_rep2007}. Comparing our results to the reference, it is evident there is only a minor loss of accuracy.
    }
    \label{fig:evoAdi}
\end{figure}

The progress of the system dynamics in the time interval $[t_0=0,t_f=4000]$ is monitored by measuring the expectation values of the energy and of the proton and electron occupation numbers. 
The energies associated with the three Hamiltonians
\begin{equation}
    E_X = \expval{\hat{H}_X}{\Psi(t)},\quad  X\in\{L,M,R\}\, ,
\end{equation}
are used to examine the time-dependent expectation value $E(t)$ associated with the Hamiltonian in Eq.~\eqref{eq:H_linear_comb}, which is shown in Fig.~\ref{fig:evoAdi} (lower panel). 
In addition, we also monitor the progress of the electron transfer dynamics by measuring the expectation values of the electron and the proton occupation numbers in their corresponding spin-orbitals, using the operators $\hat{N}_i=\acr{i} \aan{i}$ and $\hat{N}_{I}=\acr{I} \aan{I}$, respectively. The occupation numbers are given by
\begin{equation}
    n_{I/i} = \expval{\hat N_{I/i}}{\Psi(t)} \, ,
    \label{eq:n_I} 
\end{equation}
where the index $i$ labels the electronic spin-orbitals shown in Fig.~\ref{fig:malon_orbitals} (red and blue), while $I$ labels the nuclear spin-orbitals (orange). 
Of particular interest are the occupations of the three different nuclear spin-orbitals $n_L$,$n_M$,$n_R$ (\textit{i.e.}, Eq.~\eqref{eq:n_I} with $I=L,M,R$), which reach their highest values for the ground state of the corresponding Hamiltonians, $\hat H_L, \hat H_M, \hat H_R$. Their time evolution is shown in the second panel from the bottom of Fig.~\ref{fig:evoAdi}. 
Similarly, we also monitored the occupations of the first two $\alpha$ and $\beta$ electronic spin orbitals (see the second panel from the top in Fig.~\ref{fig:evoAdi}).

Of particular interest is also the measure of the state fidelity, $\mathcal{F}$, which is computed as the squared absolute overlap between the time-dependent state and the ground states of the reference Hamiltonians $\Psi_L$, $\Psi_M$, and $\Psi_R$ 
\begin{equation}\label{eq:fid_pure}
    \mathcal{F}_X\left(\ket{\Psi(t)}, \ket{\Psi_X}\right) = |\braket{\Psi(t)}{\Psi_X}|^2,\quad X\in\{L,M,R\}\, .
\end{equation}
$\mathcal{F}_X$ takes a value in the range $[0,1]$, where 1 corresponds to a maximum absolute overlap and therefore to identical (up to a global phase) states. In particular, we will use the fidelities $\mathcal{F}_M$ and $\mathcal{F}_R$ to assess the adiabaticity of our time-evolution protocol.

\begin{figure}[ht]
    \centering
    \includegraphics[width=\plotwidth]{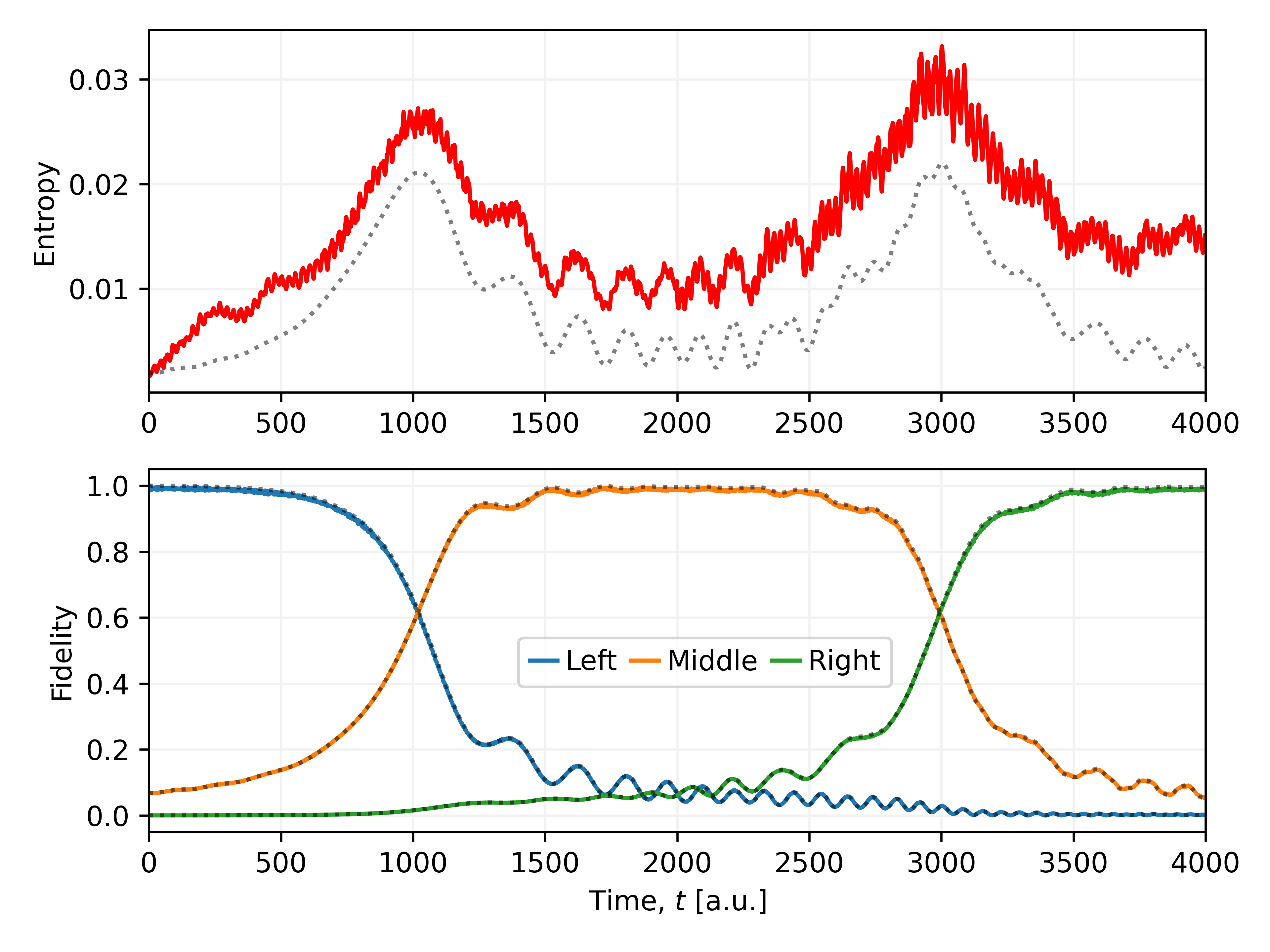}
    \caption{Nuclear-electron entanglement entropy (top) and the fidelities with respect to reference ground states in the `Left', `Middle', `Right' setups (bottom) as a function of time in the adiabatic evolution for $t_f=4000$ a.u. with $\Delta t = 1$ a.u.. The reference values, generated using RK4~\cite{num_rep2007}, are shown as black dotted lines. While the fidelities are in good agreement with reference values, the entanglement entropy acquires a significant error (note that it is orders of magnitude smaller than the fidelity).}
    \label{fig:FideAdi}
\end{figure}

\begin{figure}[ht]
    \centering
    \includegraphics[width=\plotwidth]{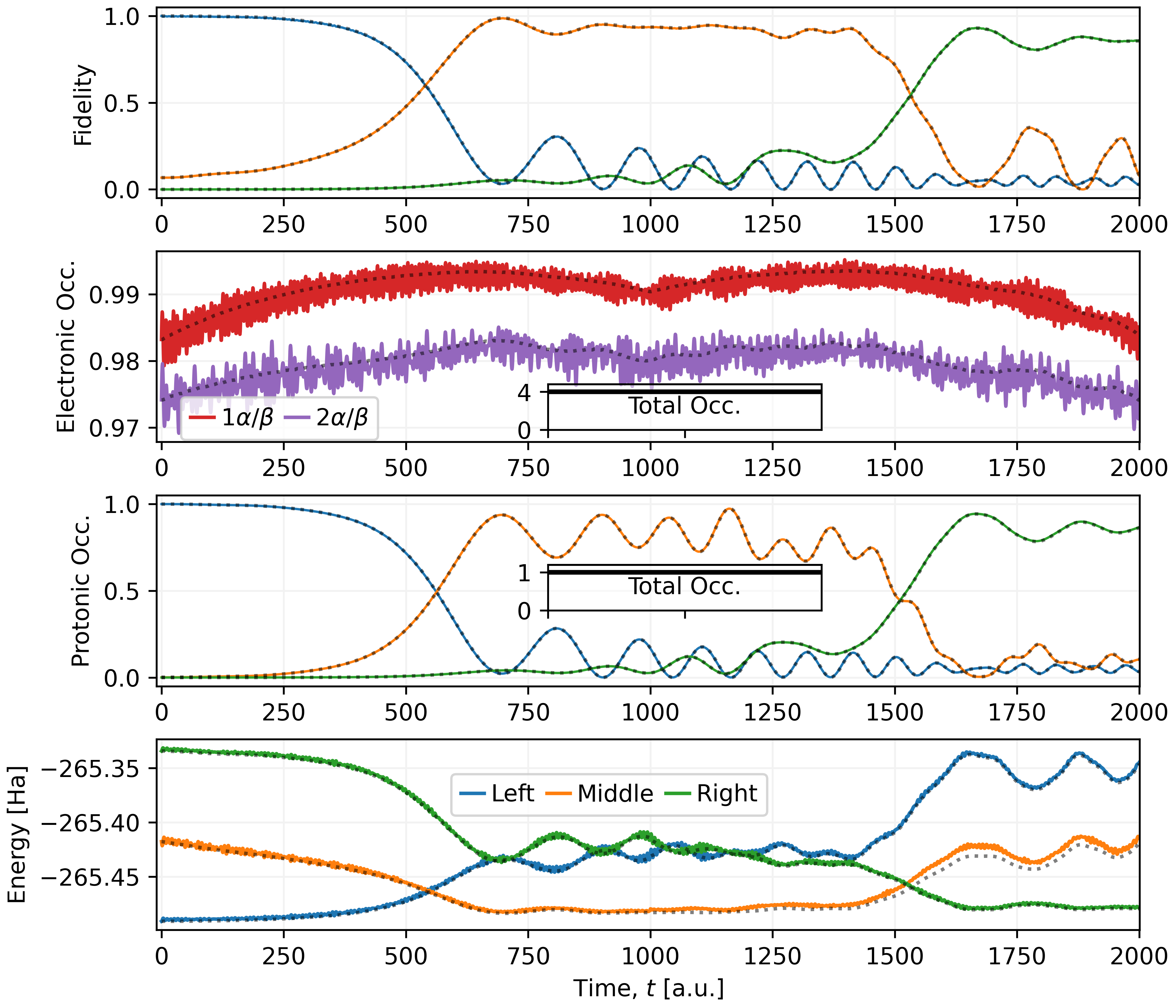}
    \caption{
    Dynamics performed for time $t_f=2000$ a.u. with $\Delta t = 0.5$ a.u.. The results are in good agreement with the reference evolution, generated using RK4~\cite{num_rep2007}, shown as black dotted lines.}
    \label{fig:evoAdiFaster}
\end{figure}

\textit{Result and Discussion.} 
In the following, we present two time evolution settings that we named for convenience `slow' (or adiabatic) and `fast' (or nonadiabatic). 
In the first case, the entire Hamiltonian evolution (Eq~\eqref{eq:H_linear_comb}) is completed in 4000 time steps with step size $\Delta t= 1.0$ a.u., which (as we will see below) is sufficient to closely approximate adiabtic evolution of the proton-electron state (see Fig.~\ref{fig:evoAdi}). In the second case, we double the speed at which we perform the dynamics, acting in a nonadiabatic, `fast', regime (see Fig.~\ref{fig:evoAdiFaster}). This is realized by halving the step size, $\Delta t = 0.5$ a.u., and keeping the same total number of steps, 4000.
In the `slow', adiabatic case, the change over time of parameters $\alpha$, $\beta$, and $\gamma$ results in a one-way transfer of the proton from the oxygen in the left position (L) to the one on the right (R) (see Fig~\ref{fig:schematic}). 
The transfer is clearly evidenced by the evolution of the nuclear occupation numbers in Figure~\ref{fig:evoAdi} (second panel from the bottom) and the fidelity plot in Figure~\ref{fig:FideAdi}. 
The occupation numbers of the nuclear spin-orbitals evolve from a set of values $(n_L=1, n_M=0, n_R=0)$ at $t=0$ a.u., transitioning to an intermediate regime dominated by $n_M \sim 1$ around time 2000 a.u., and finally converging towards the values $(n_L=0, n_M=0, n_R=1)$  at the end of the simulation. 
The time evolution of the system energy (see the bottom panel in Figure~\ref{fig:evoAdi}) inversely mirrors the changes just described for the occupation numbers. 
At the beginning of the simulation, the expectation values for $\hat H_M$ ($E_M$) and $\hat H_R$ ($E_R$) greatly exceed the corresponding ground state energies as the starting \ac{neouccsdt} state is optimized for the $\hat H_L$ ground state, see Figure~\ref{fig:evoAdi}. 
Then, as the simulation progresses towards the middle position (dominated by $\hat H_M$), $E_L$ slowly grows while $E_M$ and $E_R$ decrease. 
At the halfway mark, ($t\sim 2000$ a.u.) the $E_L$ and $E_R$ values are nearly equal, while $E_M$ closely approaches the ground state energy value of $\hat H_M$. 
This means that our protocol is capable of driving the system adiabatically from the initial state (with the proton localized to the left) to the middle configuration with the equally shared proton.  
This is also supported by the values of occupations in the middle panel of Figure~\ref{fig:evoAdi}. 
A similar description --- in reverse order --- applies to the second half of the dynamics leading to a final state with a large overlap with the ground state of the Hamiltonian $\hat H_R$. 
Concerning the electronic structure, we also observe changes in the occupation of the spin orbitals involved in the calculations (the second panel from the top of Figure~\ref{fig:evoAdi}). 
Specifically, the occupation numbers of the first two lower-lying spin orbitals increase slightly during the proton transfer process reaching a maximum at the transition barrier of the L-M and M-R half-reactions. Note that the occupations refer to both $\alpha$ and $\beta$ spin orbitals because we are in the restricted picture. It is worth stressing that for both protons and electrons, the total number of particles (integrated over all basis functions) remains conserved throughout the full reaction path and amounts to 1 and 4, respectively. 
Constant particle numbers confirm the absence of particle leakage, which is a signature of faithful implementation of the unitary time-evolution protocol. 
The high-frequency oscillations observed in the nuclear occupancy numbers correspond to the nonadiabatic part of the coherent time evolution, \textit{i.e.}, slight deviations from the instantaneous ground state of $\hat H(t)$. These can be suppressed if we further decrease the rate of change of the Hamiltonian coefficients $\alpha$, $\beta$ and $\gamma$, thus approaching the adiabatic limit. This can be seen in SI Fig.~\ref{fig:evoAdiST0001}, where the adiabatic speed (change in $\alpha$, $\beta$, and $\gamma$ per $\Delta t=1.0$ a.u.) was five times slower.

\begin{figure}[H]
    \centering
    \includegraphics[width=\plotwidth]{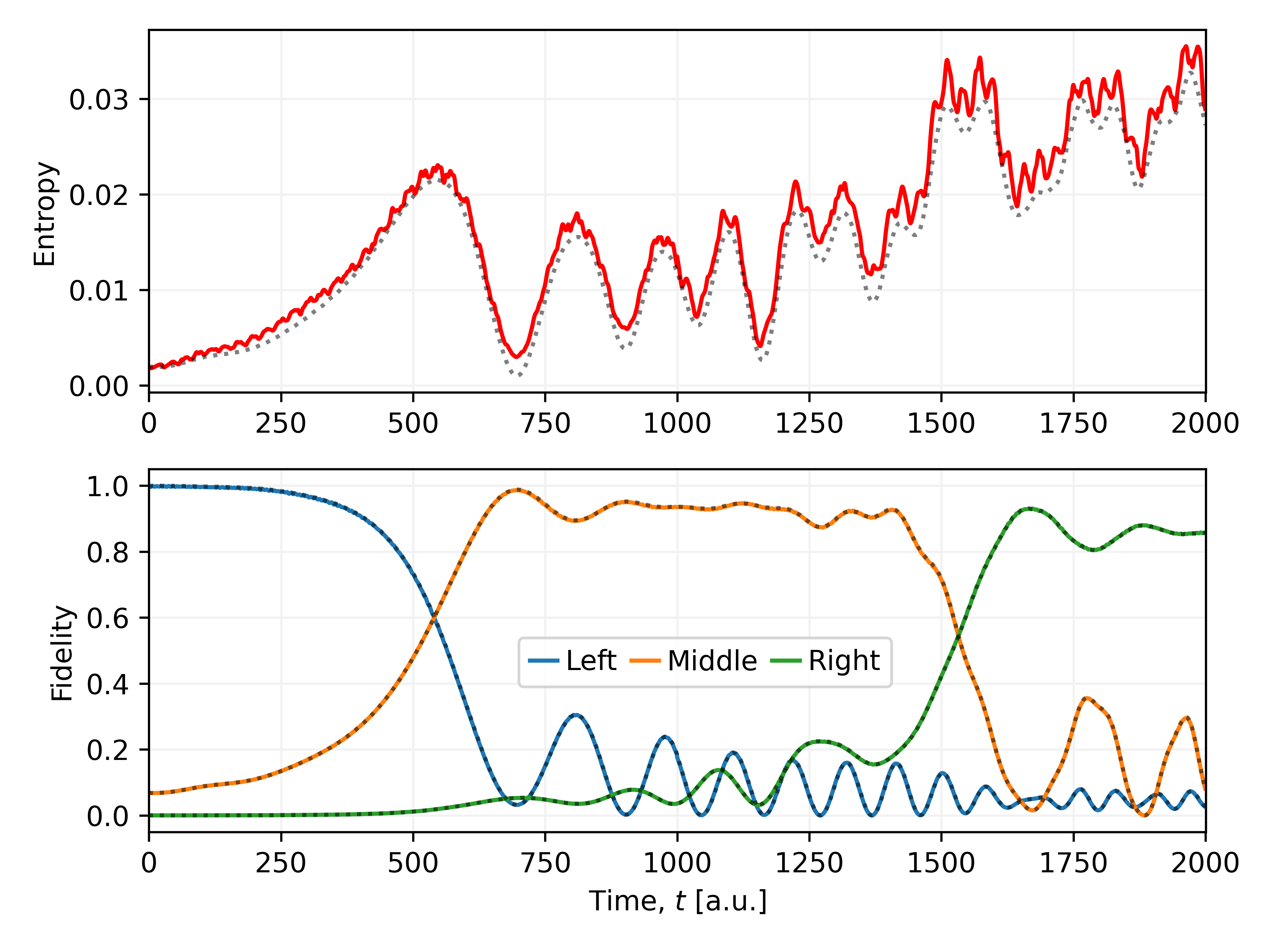}
    \caption{
    Nuclear-electron entanglement entropy (top) and the fidelities with respect to reference ground states in the `Left', `Middle', `Right' setups (bottom) as a function of time in the dynamics for $t_f=2000$ a.u. with $\Delta t = 0.5$ a.u.. The reference values, generated using RK4~\cite{num_rep2007}, are shown as black dotted lines. As a resolution of the simulation is increased, $\Delta t = 0.5$ a.u., the entanglement entropy shows better agreement with the reference values compared to results with $\Delta t = 1$ a.u..
    }
    \label{fig:FideAdiFaster}
\end{figure}

Of particular interest is the analysis of the entanglement entropy evaluated for the two subsystems, Eq.~\eqref{eq:vonNeuman}, along the reaction path. For details, see~\cite{Note1}.
In the adiabatic regime the exact entropy profile shows a bimodal, roughly symmetric, shape with the two peaks at the L-M and M-R transition points. 
As before, the oscillations are caused by nonadiabatic effects and disappear if we further slow down the dynamics (see SI Fig.~\ref{fig:FideAdi0001}). Note that while the exact profile (dotted line, Fig.~\ref{fig:FideAdi}) fully recovers the initial value, in the first-order Suzuki~\cite{suzuki1993general} approach a residual entanglement entropy is left after full transfer of the proton (red curve, Fig.~\ref{fig:FideAdi}). This deviation is attributed to the the first-order Suzuki approximation error and can be reduced by decreasing step size to $\Delta t = 0.5$ a.u.\ as shown in Fig.~\ref{fig:FideAdiFaster}. 
The fidelities with respect to the ground state wave functions corresponding to the different components of the Hamiltonian (Eq.~\eqref{eq:H_linear_comb}) are reported in the lower panel of Fig.~\ref{fig:FideAdi} and confirm the quality of the simulations (see also the fully adiabatic profiles in SI Fig.~\ref{fig:FideAdi0001}). 
Finally, it is worth noting that entropy sharply rises at the moments when the evolving state has equal fidelity first for $\mathcal{F}_L$ and $\mathcal{F}_M$ and then later for  $\mathcal{F}_M$ and $\mathcal{F}_R$. 
At these points in the evolution there is thus a maximum amount of entanglement between the protonic and electronic degrees of freedom along the adiabatic trajectory. 

In the `fast' regime, the physics is quite different. During the entire proton transfer process, the entropy of the system keeps growing (with large fluctuations) and reaches its maximal value at the end of the simulation. In contrast to the slow regime,
we observe an accumulation of entanglement due to excitations to higher energy levels along the nonadiabatic trajectory. 
Both the exact (dashed line in Fig.~\ref{fig:FideAdiFaster}) and the first-order Suzuki~\cite{suzuki1993general} dynamics confirm this behaviour.
The entanglement will persist unless a decoherence channel is introduced. To monitor this process, one would need to include bath degrees of freedom (\textit{i.e.}, the dynamics of the molecular scaffold and of the solvent) and simulate the dynamics as an open quantum system.

\textit{Conclusions.} 
In this paper, we presented a quantum computing algorithm for modelling electron-nuclear coupled dynamics in molecular systems formulated in the second quantization framework. 
In classical computation setups, nuclear and electronic degrees of freedom exponentially increase with the system size, leading to unfavourable scaling in both memory and execution time. 
On the other hand, quantum computers have the potential to solve these same problems in polynomial time and using polynomial memory. 
The quantum algorithm was tested with classical simulations leading to very accurate results and interesting insights about the proton transfer process in a realistic model of malonaldehyde. 
In particular, we showed that when the dynamics is sufficiently fast, such that the electrons cannot follow adiabatically, entanglement between electronic and nuclear degrees of freedom is generated and persists over time. 
The classical emulation of the quantum algorithm will quickly become unfeasible as the number of electronic and nuclear basis functions (and therefore qubits) exceeds about 20-30 qubits.
From this point on, only quantum computers will be able to perform the coupled electron-nuclear quantum dynamics at such large scales. 
However, the noise in state-of-the-art quantum computers currently hampers demonstrations of long-time adiabatic quantum simulations of complex systems with more than 10-20 qubits, due to the required circuit depth. 
Further developments are therefore needed for a more efficient implementation of the proposed electron-nuclear quantum dynamics scheme before reaching the fault-tolerant regime. 
In particular, we are planning to investigate more efficient encoding schemes for the time evolution operator in conjunction with error mitigation schemes~\cite{berg2022probabilistic}, as well as the possibility of applying variational time-evolution algorithms.

\section*{Acknowledgements}
This work was supported by the Hartree National Centre for Digital Innovation, a collaboration between STFC and IBM. This research was supported by funding from Horizon 2020 via the NEASQC project (grant number 951821), the Wallenberg Center for Quantum Technology (WACQT), and the NCCR MARVEL, a National Centre of Competence in Research, funded by the Swiss National Science Foundation (grant number 205602). IBM, the IBM logo, and ibm.com are trademarks of International Business Machines Corp., registered in many jurisdictions worldwide. Other product and service names might be trademarks of IBM or other companies. The current list of IBM trademarks is available at \url{https://www.ibm.com/legal/copytrade}.

\appendix

\section*{Supporting Information (SI)}

\begin{figure}[H]
    \centering
    \includegraphics[width=\plotwidth]{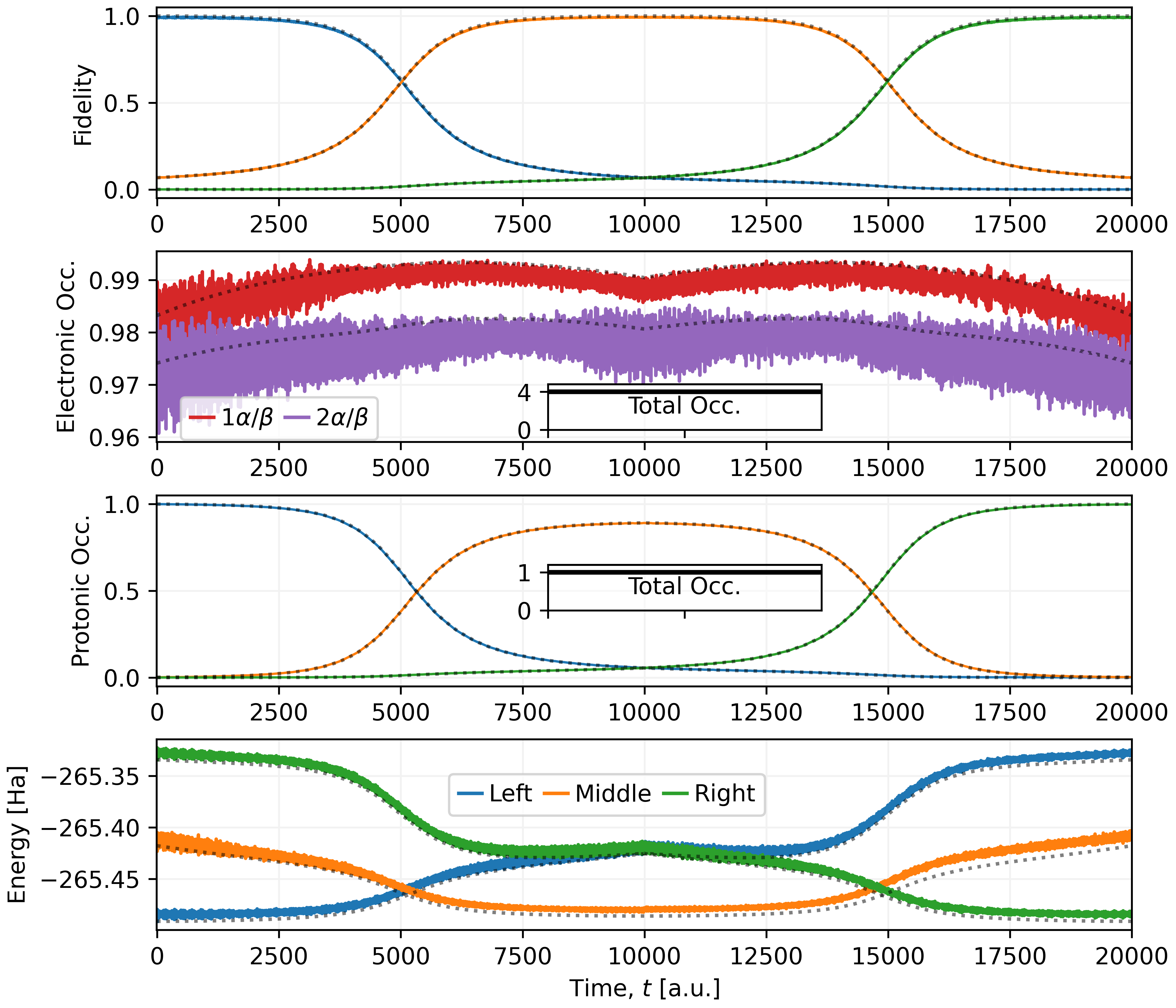}
    \caption{
    Adiabatic evolution performed for time $t_f=20000$ a.u.\ with $\Delta t = 1.0$ a.u.. There are slight deviations from the reference results which are shown as black dotted lines. The reference evolution uses the RK4 method~\cite{num_rep2007}. 
    }
    \label{fig:evoAdiST0001}
\end{figure}

\begin{figure}[H]
    \centering
    \includegraphics[width=\plotwidth]{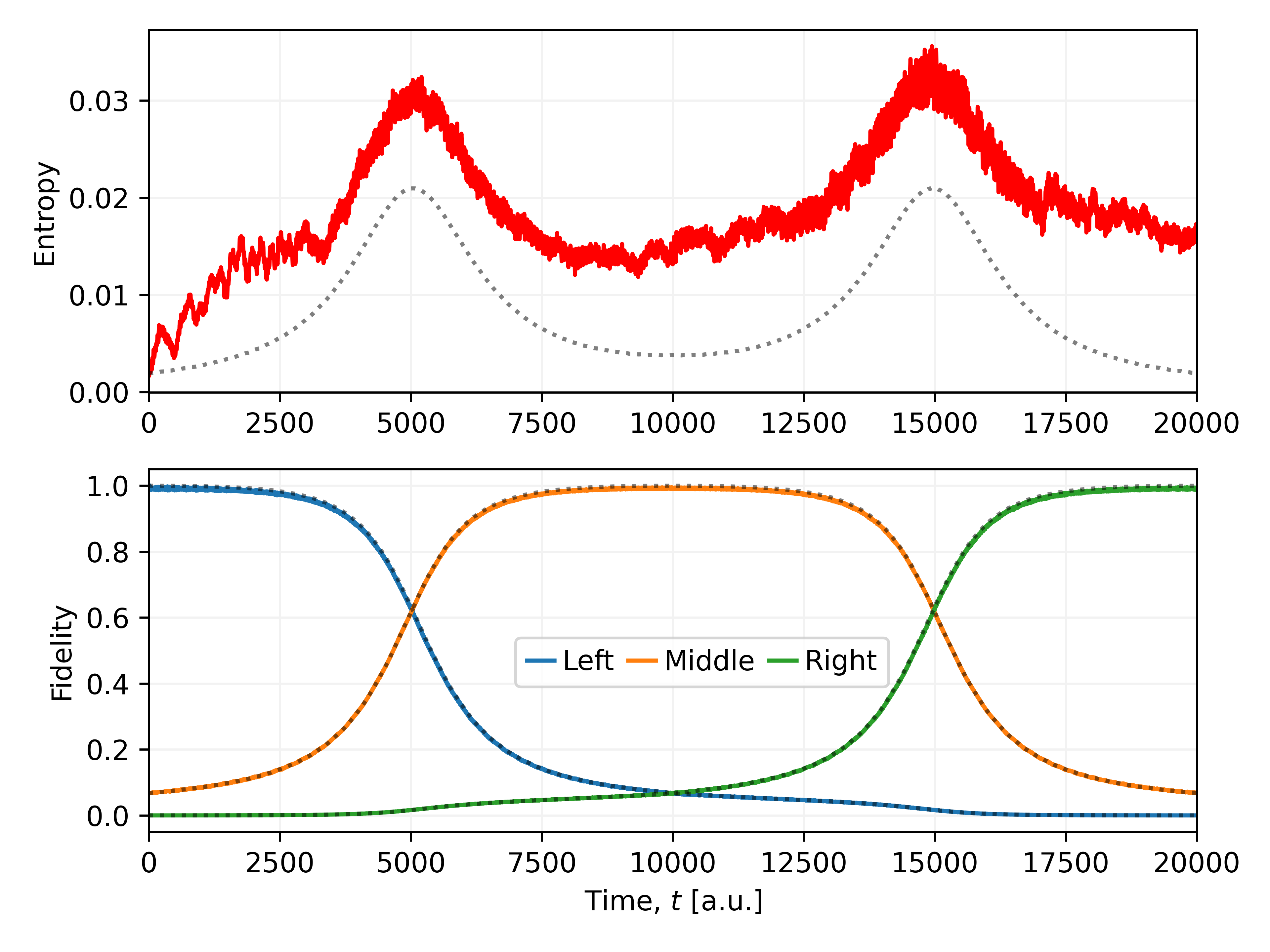}
    \caption{
    Nuclear-electron entanglement entropy (top) and the fidelities with respect to reference ground states in the `Left', `Middle', `Right' setups (bottom) as a function of time in the adiabatic evolution for $t_f=20000$ a.u.\ with $\Delta t = 1$ a.u.. The reference values, generated using RK4~\cite{num_rep2007}, are shown as black dotted lines. The values for fidelities and entanglement entropy show similar results to the shorter evolution with $t_f=4000$ a.u.\, proving that the evolution operator is stable over a long-time simulation and does not accumulate errors.
    }
    \label{fig:FideAdi0001}
\end{figure}


\begin{thebibliography}{10}

\bibitem{wild2023rate}
Robert Wild and Roland Wester.
\newblock Rate of quantum-tunnelling reaction revealed.
\newblock {\em Measurement}, 10:3, 2023.

\bibitem{schreiner2008capture}
Peter~R Schreiner, Hans~Peter Reisenauer, Frank~C Pickard~Iv, Andrew~C Simmonett, Wesley~D Allen, Edit M{\'a}tyus, and Attila~G Cs{\'a}sz{\'a}r.
\newblock Capture of hydroxymethylene and its fast disappearance through tunnelling.
\newblock {\em Nature}, 453(7197):906--909, 2008.

\bibitem{schreiner2011methylhydroxycarbene}
Peter~R Schreiner, Hans~Peter Reisenauer, David Ley, Dennis Gerbig, Chia-Hua Wu, and Wesley~D Allen.
\newblock Methylhydroxycarbene: Tunneling control of a chemical reaction.
\newblock {\em Science}, 332(6035):1300--1303, 2011.

\bibitem{klinman2013hydrogen}
Judith~P Klinman and Amnon Kohen.
\newblock Hydrogen tunneling links protein dynamics to enzyme catalysis.
\newblock {\em Annual review of biochemistry}, 82:471--496, 2013.

\bibitem{cha1989hydrogen}
Yuan Cha, Christopher~J Murray, and Judith~P Klinman.
\newblock Hydrogen tunneling in enzyme reactions.
\newblock {\em Science}, 243(4896):1325--1330, 1989.

\bibitem{nagel200921st}
Zachary~D Nagel and Judith~P Klinman.
\newblock A 21st century revisionist's view at a turning point in enzymology.
\newblock {\em Nature chemical biology}, 5(8):543--550, 2009.

\bibitem{bavnacky1981dynamics}
Pavol Ba{\v{n}}ack{\`y}.
\newblock Dynamics of proton transfer and enzymatic activity.
\newblock {\em Biophysical Chemistry}, 13(1):39--47, 1981.

\bibitem{basran1999enzymatic}
Jaswir Basran, Michael~J Sutcliffe, and Nigel~S Scrutton.
\newblock Enzymatic h-transfer requires vibration-driven extreme tunneling.
\newblock {\em Biochemistry}, 38(10):3218--3222, 1999.

\bibitem{slocombe2022open}
Louie Slocombe, Marco Sacchi, and Jim Al-Khalili.
\newblock An open quantum systems approach to proton tunnelling in dna.
\newblock {\em Communications Physics}, 5(1):109, 2022.

\bibitem{verma2021scaling}
Prakash Verma, Lee Huntington, Marc~P Coons, Yukio Kawashima, Takeshi Yamazaki, and Arman Zaribafiyan.
\newblock Scaling up electronic structure calculations on quantum computers: The frozen natural orbital based method of increments.
\newblock {\em The Journal of Chemical Physics}, 155(3):034110, 2021.

\bibitem{kovy2023}
Arseny Kovyrshin, Mårten Skogh, Anders Broo, Stefano Mensa, Emre Sahin, Jason Crain, and Ivano Tavernelli.
\newblock A quantum computing implementation of nuclear-electronic orbital (neo) theory: Towards an exact pre-born-oppenheimer formulation of molecular quantum systems.
\newblock {\em arXiv:2302.07814}, 2023.


\bibitem{webb2002}
Hammes-Schiffer~{S.} Webb~{S. P.}, Iordanov~{T.}
\newblock Multiconfigurational nuclear-electronic orbital approach: Incorporation of nuclear quantum effects in electronic structure calculations.
\newblock {\em J. Chem. Phys.}, 117:4106--4118, 2002.

\bibitem{zhao2020real}
Luning Zhao, Zhen Tao, Fabijan Pavošević, Andrew Wildman, Sharon Hammes-Schiffer, and Xiaosong Li.
\newblock Real-time time-dependent nuclear- electronic orbital approach: dynamics beyond the born--oppenheimer approximation.
\newblock {\em The Journal of Physical Chemistry Letters}, 11(10):4052--4058, 2020.

\bibitem{pavošević2020multicomponent}
Fabijan Pavošević, Tanner Culpitt, and Sharon Hammes-Schiffer.
\newblock Multicomponent quantum chemistry: Integrating electronic and nuclear quantum effects via the nuclear--electronic orbital method.
\newblock {\em Chemical reviews}, 120(9):4222--4253, 2020.

\bibitem{coutinho2004ground}
Maur{\i}cio~D Coutinho-Neto, Alexandra Viel, and Uwe Manthe.
\newblock The ground state tunneling splitting of malonaldehyde: Accurate full dimensional quantum dynamics calculations.
\newblock {\em The Journal of chemical physics}, 121(19):9207--9210, 2004.

\bibitem{yu2022nonadiabatic}
Qi~Yu, Saswata Roy, and Sharon Hammes-Schiffer.
\newblock Nonadiabatic dynamics of hydrogen tunneling with nuclear-electronic orbital multistate density functional theory.
\newblock {\em Journal of Chemical Theory and Computation}, 18(12):7132--7141, 2022.

\bibitem{suzuki1993general}
Masuo Suzuki.
\newblock General decomposition theory of ordered exponentials.
\newblock {\em Proceedings of the Japan Academy, Series B}, 69(7):161--166, 1993.

\bibitem{yuan2019}
Xiao Yuan, Suguru Endo, Qi~Zhao, Ying Li, and Simon~C. Benjamin.
\newblock Theory of variational quantum simulation.
\newblock {\em {Quantum}}, 3:191, 2019.

\bibitem{mcar2019}
Sam McArdle, Tyson Jones, Suguru Endo, Ying Li, Simon Benjamin, and Xiao Yuan.
\newblock Variational ansatz-based quantum simulation of imaginary time evolution.
\newblock {\em Npj Quantum Inf.}, 5:75, 2019.

\bibitem{num_rep2007}
{William H.} Press, {Saul A.} Teukolsky, {William T.} Vetterling, and {Brian  P.} Flannery.
\newblock {\em Numerical Recipes 3rd Edition: The Art of Scientific Computing}.
\newblock Cambridge University Press, New York, NY, USA, 3 edition, 2007.

\bibitem{Note1}
For the entanglement entropy to show the symmetric pattern for the bond
  breaking and dissociation with the right and left oxygen atoms, one needs to
  align the phase signature for the orbitals of the ``Middle'' and ``Right''
  setups. We perform the similarity transformation for the ``Middle'' Hamiltonian and wave function at the midpoint of the simulation --- equivalent to flipping the phase signature on orbitals of the middle setup, see Fig.~\ref {fig:malon_orbitals}.

\bibitem{berg2022probabilistic}
Ewout van~den Berg, Zlatko~K. Minev, Abhinav Kandala, and Kristan Temme.
\newblock Probabilistic error cancellation with sparse pauli-lindblad models on noisy quantum processors, 2022.
\newblock {\em arXiv:2201.09866}, 2022.

\end{thebibliography}

\end{document}